\documentclass[aip,reprint,numerical,graphicx]{revtex4-1}

\usepackage{graphicx}
\usepackage{dcolumn}

\draft 

\begin{document}


\title{On the limits of coercivity in permanent magnets} 



\author{J. Fischbacher}
\author{A. Kovacs}
\author{H. Oezelt}
\author{M. Gusenbauer}
\author{T. Schrefl}
\email[Electronic mail: ]{tschrefl@gmail.com}
\affiliation{Center for Integrated Sensor Systems, Danube University Krems, 2700 Wiener Neustadt, Austria}

\author{L. Exl}
\affiliation{Faculty of Mathematics, Vienna University, 1090 Vienna, Austria}

\author{D. Givord}
\author{N. M. Dempsey}
\affiliation{CNRS, Institut N\'{e}el, 25 rue des Martyrs, 38042 Grenoble, France}

\author{G. Zimanyi}
\affiliation{Department of Physics, University of California, Davis, California 95616, USA}

\author{M. Winklhofer}
\affiliation{Carl von Ossietzky University of Oldenburg, 26129 Oldenburg, Germany}

\author{G. Hrkac}
\affiliation{College of Engineering, University of Exeter, Exeter, EX4 4QF, UK}

\author{R. Chantrell}
\affiliation{Department of Physics, University of York, York, YO10 5DD, UK}

\author{N. Sakuma}
\author{M. Yano}
\author{A. Kato}
\author{T. Shoji}
\affiliation{Toyota Motor Corporation, 1200 Mishuku, Susono, Shizuoka 410-1193, Japan}
\affiliation{Technology Research Association of Magnetic Materials for High-efficiency Motors (Mag-HEM) Higashifuji-Branch, 1200 Mishuku, Susono, Shizuoka 410-1193, Japan}

\author{A. Manabe}
\affiliation{Technology Research Association of Magnetic Materials for High-efficiency Motors (Mag-HEM) Higashifuji-Branch, 1200 Mishuku, Susono, Shizuoka 410-1193, Japan}


\date{\today}

\begin{abstract}
The maximum coercivity that can be achieved for a given hard magnetic alloy is estimated by computing the energy barrier for the nucleation of a reversed domain in an idealized microstructure without any structural defects and without any soft magnetic secondary phases. For Sm$_{1-z}$Zr$_z$(Fe$_{1-y}$Co$_y$)$_{12-x}$Ti$_x$ based alloys, which are considered an alternative to Nd$_2$Fe$_{14}$B magnets with lower rare-earth content, the coercive field of a small magnetic cube is reduced to 60 percent of the anisotropy field at room temperature and to 50 percent of the anisotropy field at elevated temperature (473K). This decrease of the coercive field is caused by misorientation, demagnetizing fields and thermal fluctuations.
\end{abstract}


\maketitle 


Permanent magnets are an important material for energy conversion in modern technologies. Wind power as well as hybrid and electric vehicles require high performance permanent magnets. In motor applications the magnet should retain a high magnetization and coercive field at an operating temperature around 450 K. At this temperature the magnetization and the anisotropy field of Sm$_{1-z}$Zr$_z$(Fe$_{1-y}$Co$_y$)$_{12-x}$Ti$_x$ are higher than those of Nd$_{2}$Fe$_{14}$B. \cite{kuno2016sm} In addition, the rare earth to transition metal ratio of the SmFe$_{12}$ based magnets is lower. Therefore, magnets based on this phase are considered as a possible alternative to Nd$_2$Fe$_{14}$B magnets. \cite{hirosawa2017perspectives}
At high temperature, thermal fluctuations may reduce the coercive field. In this work, we numerically compute the reduction of coercivity by thermal fluctuations in Sm$_{1-z}$Zr$_z$(Fe$_{1-y}$Co$_y$)$_{12-x}$Ti$_x$. For comparison, we also include results for Nd$_{2}$Fe$_{14}$B. The letter is organized as follows. We first review the different effects that reduce the coercive field in permanent magnets. Then we present a numerical method for the computation of the coercive field including thermal fluctuations, which is based on finite element micromagnetics. We introduce the concept of the activation volume which is widely used in the experimental analysis of coercivity in permanent magnets. Then we present numerical results for Nd$_{2}$Fe$_{14}$B and Sm$_{1-z}$Zr$_z$(Fe$_{1-y}$Co$_y$)$_{12-x}$Ti$_x$. 

Besides thermal fluctuations, several other effects reduce the coercive field of modern permanent magnets. Kronm{\"u}ller et al. \cite{kronmuller1988analysis} refer to the difference between the anisotropy field of a magnet and its coercive field as a discrepancy from theory. Aharoni \cite{aharoni1960reduction} predicted that the coercive field of a hard magnet decreases with increasing width of surface defects with zero anisotropy. The corresponding minimum coercive field is $1/4$ of the anisotropy field which is reached for a defect width greater than 5$\sqrt{A/K}$, where $A$ is the exchange constant and $K$ is the anisotropy constant. Even smaller coercive fields may occur if the anisotropy increases gradually from zero to its maximum value as shown by  Becker and D{\"o}ring \cite{Becker1939} and Hagedorn \cite{hagedorn1970analysis}. 

In addition to defects, local demagnetizing fields reduce the coercivity of permanent magnets. Gr{\"o}nefeld and Kronm{\"u}ller \cite{gronefeld1989calculation} show that the local demagnetizing field may reach values of the order of the saturation magnetization, $M_{\mathrm s}$, near the edges of a hard magnetic grain. The total field which is essential for the switching of a grain is the sum of the local demagnetizing field and the external field. Therefore, the local demagnetizing field leads to a further reduction of coercivity. 

A further reduction of the coercive field as compared to the ideal nucleation field, $H_{\mathrm N} = 2K/(\mu_0 M_{\mathrm s})$, may result from dynamic effects  \cite{leineweber1999dynamics}. When the external field or the internal effective field is changing at a rate much faster than the energy dissipation in the system, the system cannot follow fast changes in the energy landscape and thus does not reach the nearest metastable state. Instead a path through the energy landscape that brings the system into a reversed magnetic state may be taken.  Leineweber and Kronm{\"u}ller \cite{leineweber2000influence} show that dynamic effects can reduce the ideal nucleation field by up to 20 percent.

In this work we focus on thermal fluctuations and calculate the reduction of coercivity caused by these fluctuations. Magnetization reversal in a permanent magnet is the process by which an external field creates a reversed nucleus near structural defects. Thermal fluctuations assist the formation of the reversed nucleus and thus reduce the coercive field. The formation of the nucleus is associated with an energy barrier. Before magnetization reversal the system is in a local energy minimum. With increasing external field, the energy barrier that separates the local minimum from the reversed magnetic state decreases. \cite{schabes1991micromagnetic} Taking into account thermal activation the system can overcome an energy barrier, $E$, within a time $\tau = \tau_0 \exp⁡ \left(E/\left(k_{B}T\right)\right)$.\cite{Becker1939} Here $k_\mathrm{B} = 1.38 \times 10^{-23}$~J/K is the Boltzmann constant. The time constant $\tau_0$ is the inverse of the attempt frequency $f_0$. Often it is assumed that the magnet can overcome an energy barrier of $25k_\mathrm{B}T$ within the time $\tau = 1$~s which gives an attempt frequency of $f_0 = 7.2 \times10^{10}$~s$^{-1}$.\cite{gaunt1976magnetic} Then the coercive field is the critical value of the external field, $H$, at which the energy barrier $E(H)$ reaches $25k_\mathrm{B}T$.   

Using numerical micromagnetics, we compute the energy barrier as a function of the applied field. We discretize the magnet's microstructure with tetrahedral finite elements. Minimizing the energy  for varying external field gives the magnetic states along the demagnetization curve. For energy minimization we apply the non-linear conjugate gradient method as described by Fischbacher and co-workers \cite{fischbacher2017nonlinear}. The coercive field obtained from the computation of the demagnetization curve is $H_0$. This is the field at which the energy barrier is zero. We now want to compute the energy barrier for a field $H < H_{0}$. We apply the string method \cite{zhang2016recent} in order to compute the minimum energy path that connects the local minimum at field $H$ with the reversed magnetic state. A path is called a minimum energy path, if for any point along the path the gradient of the energy is parallel to the path. In other words; the component of the energy gradient
normal to the path is zero. The magnetization configurations along the path are described by images. Each image is a replica of the total system. The minimum energy path over a saddle point is found iteratively. A single iteration step consists of two moves. First each image is relaxed \cite{samanta2013optimization} by applying a few steps of the conjugate gradient method, then the images are moved along the path so that the distance between the images is constant. We use an energy weighted distance and truncate the path \cite{carilli2015truncation} so that there are more images next to the saddle point. We repeat the computation of the minimum energy path for different applied fields  and obtain $E(H)$. We compute $H_{\mathrm c}(T)$ by the intersection of the $E(H)$ curve with the line $E = 25k_\mathrm{B}T$ (see Fig. 1).

Path finding algorithms are well established both in chemical physics as well as in micromagnetics.\cite{zhang2016recent} As shown in Fig. 1 the applied algorithms are self-consistent. The switching field obtained by a classical micromagnetic method is equal to the critical field at which the computed energy barrier vanishes.  Please note that the computation of the demagnetization curve by energy minimization \cite{brown1963micromagnetics} and the computation of the minimum energy path uses the same computational grid and the same numerical minimization algorithm. Thermal fluctuations at the atomistic level are taken into account by using temperature dependent intrinsic magnetic properties such as $M_\mathrm{s}(T)$, $K(T)$, and $A(T)$. 

The above numerical scheme takes into account thermal activation over finite energy barriers. Skomski et al. \cite{skomski2013finite} reported another mechanism of coercivity reduction by thermal fluctuations. Spin waves interact with small soft magnetic structural defects which in turn cause a reduction of coercivity. The corresponding change in coercivity was found to be less than one percent. In our analysis this effect is not taken into account.

We can express the coercive field as
\begin{equation}
\label{eq_withHf}
H_{\mathrm c}= \alpha H_{\mathrm N}-N_\mathrm{eff}M_{\mathrm s}-H_{\mathrm f}. 
\end{equation}
Expression (\ref{eq_withHf}) is reminiscent of the micromagnetic equation\cite{kronmuller1988analysis} often used to analyze the temperature dependence of coercivity in hard magnets. The coefficient  $\alpha$ expresses the reduction in coercivity due to defects, misorientation, and intergrain exchange interactions.\cite{kronmueller1996} The microstructural parameter $N_\mathrm{eff}$ is related to the effect of the local demagnetization field near sharp edges and corners of the microstructure. The fluctuation field $H_\mathrm{f}$ gives the reduction of the coercive field by thermal fluctuations. \cite{givord1988coercivity} In this work, we will quantify the different effects that reduce the coercivity according to (1). In particular we are interested in the limits of coercivity. By computing $\alpha$, $N_\mathrm{eff}$, and $H_\mathrm{f}$ for a perfect hard magnetic particle without any defect we can estimate the maximum possible coercive field for a given magnetic material and microstructure. This is especially important considering the current effort to search for new hard magnetic phases with reduced rare-earth content \cite{hirosawa2017perspectives}. In addition, one might take into account the thermal fluctuation field to know how much magnetic anisotropy is enough for a permanent magnet \cite{skomski2016magnetic}. 
The coercive field which would be measured in the absence of thermal activation is $H_0 = \alpha H_{\mathrm N} - N_\mathrm{eff}M_{\mathrm s}$. 
 
The height of the energy barrier as a function of field, $E(H)$, can be derived from viscosity measurements, series expansion, or micromagnetic simulations. N{\'e}el \cite{neel1950theorie} derived a series expansion of the form $E = c \left( H_0 - H \right) ^m$ to describe the field dependence of the energy barrier, where $c$ is a constant. Analyzing the micromagnetic free energy, Skomski et al. \cite{skomski2006micromagnetic} showed that physically reasonable exponents are $m = 3/2$ and $m = 2$. 
The numerical algorithm presented above does not make any prior assumption on how the energy barrier changes with the field. Instead, we compute $E(H)$ for a finite element model of a magnetic material numerically. For the analysis of experimental data, the energy barrier is often expressed by a linear approximation $E(H) = v\mu_0 M_\mathrm{s}(H_0 - H)$.\cite{skomski2008simple} The activation volume $v$ is not necessarily related to a physical volume. Solving $E(H) = 25 k_\mathrm{B}T$ for $H$ gives the coercive field. Thus, we can write (\ref{eq_withHf}) as \cite{kronmuller2003micromagnetism} 
\begin{equation}
\label{eq_Hc}
H_{\mathrm c}= \alpha H_{\mathrm N}-N_\mathrm{eff}M_{\mathrm s}-\frac{25 k_\mathrm{B}T}{v\mu_0 M_\mathrm{s}}. 
\end{equation} 
The last term in (\ref{eq_Hc}) is proportional to the magnetic viscosity coefficient \cite{street1956comparison,wohlfarth1984coefficient} 
$S_\mathrm{v} = {k_\mathrm{B}T}/({v\mu_0 M_\mathrm{s}})$,
which can be measured experimentally. Traditionally, equations of form (\ref{eq_Hc}) have been used to analyze the temperature dependence of the coercivity. \cite{becher1997magnetic,villas1998magnetic}
 
The viscosity coefficient can be written as $S_v=-k_{B}T/ ( \partial E / \partial H )$.\cite{gaunt1976magnetic} Thus, we can define the activation volume as
\begin{equation}
\label{eq_v}
v= - \frac{1}{\mu_{0}M_\mathrm{s}} \frac{\partial E}{\partial H}.  
\end{equation} 
In this work, we will use (\ref{eq_v}) to compute the activation volume, whereby $E(H)$ is computed by finite element micromagnetic simulations.

From the comparison of the numerical results with equation (\ref{eq_withHf}) we can numerically determine the microstructural parameters $\alpha$, $N_\mathrm{eff}$, and the fluctuation field $H_\mathrm{f}$:
\begin{enumerate}
\setlength{\itemsep}{-3pt} 
\item We compute the demagnetizing curve but we switch off the demagnetizing effects by neglecting the magnetostatic self-energy in the total energy. This gives $H_0^*= \alpha H_\mathrm{N}$ and we can derive $\alpha = H_0^*/H_\mathrm{N}$. 
\item We compute the demagnetizing curve taking into account the magnetostatic energy term. This gives $H_0 =\alpha H_\mathrm{N} - N_\mathrm{eff}M_\mathrm{s} = H_0^*  - N_\mathrm{eff}M_\mathrm{s}$  and we compute $N_\mathrm{eff} = (H_0^* - H_0)/M_\mathrm{s}$. 
\item We compute the coercive field including thermal activation by $E(H_\mathrm{c}) = 25 k_\mathrm{B}T$. The fluctuation field, $H_\mathrm{f} = H_0 - H_\mathrm{c}$, represents the reduction in coercivity due to thermal activation effects.
\end{enumerate}

%
\begin{table}
\caption{\label{tab_mat} Intrinsic magnetic properties used for the simulations. The table gives the anisotropy constant $K$(MJ/m$^3$), the saturation magnetization $\mu_0M_{\mathrm{s}}(\mathrm T)$, and the exchange constant $A$(pJ/m) for different temperatures $T$(K). For Nd$_2$Fe$_{14}$B the material properties are taken from Hock \cite{hock1988zuchtung} and Durst and Kronm{\"u}ller \cite{durst1986determination}. For Sm$_{1-z}$Zr$_z$(Fe$_{1-y}$Co$_y$)$_{12-x}$Ti$_x$ compounds the material properties are taken form Kuno et al. \cite{kuno2016sm}. The exchange constant is estimated.}
\begin{tabular}{l c c c c c}
\hline \hline
Material & $T$	& $\mu_0 M_\mathrm{s}$	& $K$ &	 $A$ \\ \hline
Nd$_2$Fe$_{14}$B & 300 & 1.61 & 4.30  &  7.7 \\
Nd$_2$Fe$_{14}$B & 450 & 1.29 & 2.09  &     4.89 \\
SmFe$_{11}$Ti     & 300 & 1.26 & 5.17  & 10 \\
Sm(Fe$_{0.75}$Co$_{0.25}$)$_{11}$Ti & 300 & 1.42 & 4.67  & 10 \\
Sm(Fe$_{0.75}$Co$_{0.25}$)$_{11.5}$Ti$_{0.5}$ & 300 & 1.58 & 4.57  & 10 \\
(Sm$_{0.8}$Zr$_{0.2}$)(Fe$_{0.75}$Co$_{0.25}$)$_{11.5}$Ti$_{0.5}$ & 300 & 1.63 & 4.81 & 10 \\
SmFe$_{11}$Ti & 473 & 1.02 & 2.80 & 6.5 \\
Sm(Fe$_{0.75}$Co$_{0.25}$)$_{11}$Ti & 473 & 1.28 & 2.54 & 8.1 \\
Sm(Fe$_{0.75}$Co$_{0.25}$)$_{11.5}$Ti$_{0.5}$ & 473 & 1.45 & 2.61 & 8.4 \\
(Sm$_{0.8}$Zr$_{0.2}$)(Fe$_{0.75}$Co$_{0.25}$)$_{11.5}$Ti$_{0.5}$ & 473 & 1.50 & 2.79 & 8.4 \\ \hline  \hline
\end{tabular}
\end{table}

We are particularly interested in the limits of coercivity for a given magnetic material. Therefore, we apply the above procedure for a perfect, nano-sized hard magnetic cube without any defects. The edge length of the cube is 40 nm. However, we apply the magnetic field one degree off the easy axis which is parallel to one edge of the cube. First, we apply the method for Nd$_2$Fe$_{14}$B. Then we will show the limits of coercivity for Sm$_{1-z}$Zr$_z$(Fe$_{1-y}$Co$_y$)$_{12-x}$Ti$_x$ magnets. Table \ref{tab_mat} gives the intrinsic magnetic properties used for the simulations. For the simulation, the mesh size was 1.5 nm. Without soft magnetic defects the numerically calculated reversal field computed without magnetostatic interactions corresponds to an analytic switching field estimated by Stoner and Wohlfarth \cite{stoner1948mechanism}, $H_0^* = f(\psi_0)H_\mathrm{N}$. Here $\psi_0$ denotes the angle between the applied field and the negative anisotropy direction and $f(\psi_0)= \{\cos^{2/3}⁡(\psi _{0})+\sin^{2/3}⁡(\psi _{0})\}^{-3/2}$. \cite{kronmuller1987angular}
The agreement between the finite element results without the magnetostatic energy term and the Stoner-Wohlfarth switching field was already shown previously. \cite{fischbacher2017nonlinear} For Nd$_2$Fe$_{14}$B at 300 K we obtain $\mu_0 H_0^* = 6.09$~T. The self-demagnetizing field reduces the coercive field to $\mu_0 H_0 = 5.29$~T. Finally, with thermal fluctuations the coercive field is $\mu_0 H_\mathrm{c} = 3.94$~T. Therefore, we can conclude that in Nd$_2$Fe$_{14}$B the maximum possible coercive field of a cubic grain is only 60 percent of the ideal nucleation field $H_{\mathrm N}$. The values of $\alpha$, $N_\mathrm{eff}$, $\mu_0 H_\mathrm{f}$, and $\mu_0 S_\mathrm{v}$ are 0.91, 0.5, 1.35 T, and 0.054 T, respectively. 

%
%
\begin{figure}
\includegraphics{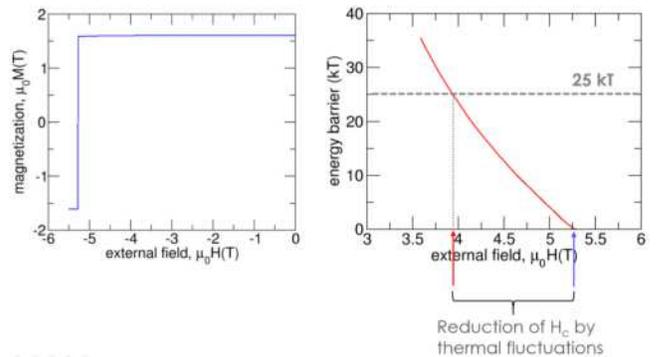}
\caption{\label{fig_demag} Left: Computed demagnetization curve for a Nd$_2$Fe$_{14}$B cube at $T = 300$~K with an edge length of 40 nm. Right: Energy barrier as a function of the external field. At the coercive field the energy barrier crosses the $25k_\mathrm{B}T$ line.}
\end{figure}

\begin{figure}
\includegraphics{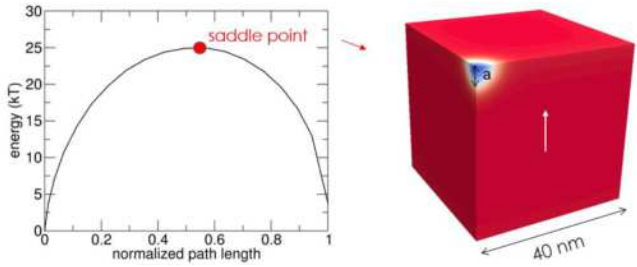}
\caption{\label{fig_nucleus} Left: Minimum energy path for a Nd$_2$Fe$_{14}$B cube at $T = 300$~K with an edge length of 40 nm. Right: Magnetization configuration of the saddle point with a reversed nucleus of size $a$.}
\end{figure}

Fig. \ref{fig_demag} gives the computed demagnetizing curve for the Nd$_2$Fe$_{14}$B cube and the energy barrier as a function of the external field computed with the intrinsic magnetic properties at $T= 300$~K. Static energy minimization for decreasing external field gives a switching field of $\mu_0 H_0 = 5.29$~T. This is exactly the field at which the energy barrier reaches zero. The reduction of coercivity owing to thermal fluctuations is 25 percent. Using (\ref{eq_v}) we can compute the activation volume, $v = (4.38$~nm)$^3$, from the slope of the $E(H)$ curve. The activation volume can be compared with the domain wall width, $ \delta = \pi \sqrt{A/K}$, which is 4.2~nm, giving $v=1.12\delta ^{3}$.\cite{givord1987magnetic} 
Fig. \ref{fig_nucleus} gives the minimum energy path and the magnetization configuration at the saddle point of the energy landscape. At the saddle point a small nucleus, which has an extension $a$, is formed. Interestingly, the volume of the reversed nucleus, $\left( 1/8 \right)  \left( 4 \pi a^{3}/3 \right)$, roughly corresponds to the activation volume $v$ as given by (\ref{eq_v}). For the small perfect cube the computed coercivity, viscosity coefficient, and the activation volume are higher than experimental values found in Nd$_2$Fe$_{14}$B based magnets.

\begin{figure}
\includegraphics{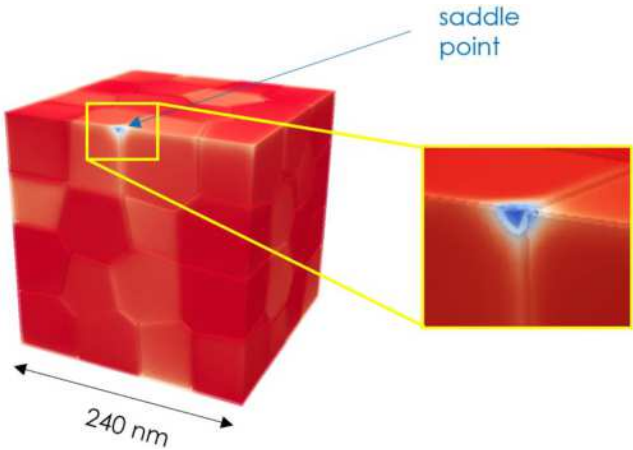}
\caption{\label{fig_grains} Saddle point of the energy for thermally assisted reversal of a multigrain Nd$_2$Fe$_{14}$B magnet. The reversed nucleus is formed at the grain boundary near the outer edge of the magnet.}
\end{figure}

For comparison with experiments we performed a similar simulation of a granular Nd$_2$Fe$_{14}$B ensemble consisting of 64 polyhedral grains with an average grain size of 60 nm. We generated the grain structure from a centroid Voronoi tessellation, using the software tool Neper \cite{quey2011large}. The grains of the Nd$_2$Fe$_{14}$B model system were separated by a weakly ferromagnetic grain boundary phase with $\mu_0 M_\mathrm{s} = 0.5$~T. The thickness of the grain boundary phase was approximately 3 nm. Grain boundaries in hot deformed Nd$_2$Fe$_{14}$B magnets were found to contain up to 55 at  $\% $ Fe. \cite{sepehri2013high} . The average misorientation angle of the grains was 15 degrees. For this magnet the values for the coercive field without magnetostatic interactions $\mu_0 H_0^*$, the intrinsic coercivity $\mu_0 H_0$, and the coercivity computed with thermal activation taken into account $\mu_0 H_\mathrm{c}$ were 3.24~T, 2.88~T, and 2.64~T, respectively. The resulting values of $\alpha$, $N_\mathrm{eff}$, $\mu_0 H_\mathrm{f}$ were 0.48, 0.22, and 0.24~T, respectively. The reduction of coercivity owing to thermal fluctuations is 8 percent. The computed viscosity coefficient $\mu_0 S_\mathrm{v}= 0.0094$~T and the computed activation volume $v = (7.9$~nm$)^3$ are very close to values measured by  Villas-Boas et al. \cite{villas1998magnetic} for a mechanically alloyed Nd$_{15.5}$Dy$_{2.5}$Fe$_{65}$Co$_{10}$Ga$_{0.75}$B$_{6.25}$ magnet at room temperature. Fig. \ref{fig_grains} shows the saddle point configuration computed from the minimum energy path. The reversed nucleus is formed in the grain boundary near the edge of the magnet. This is the location where the demagnetizing fields are the strongest.

\begin{figure}
\includegraphics{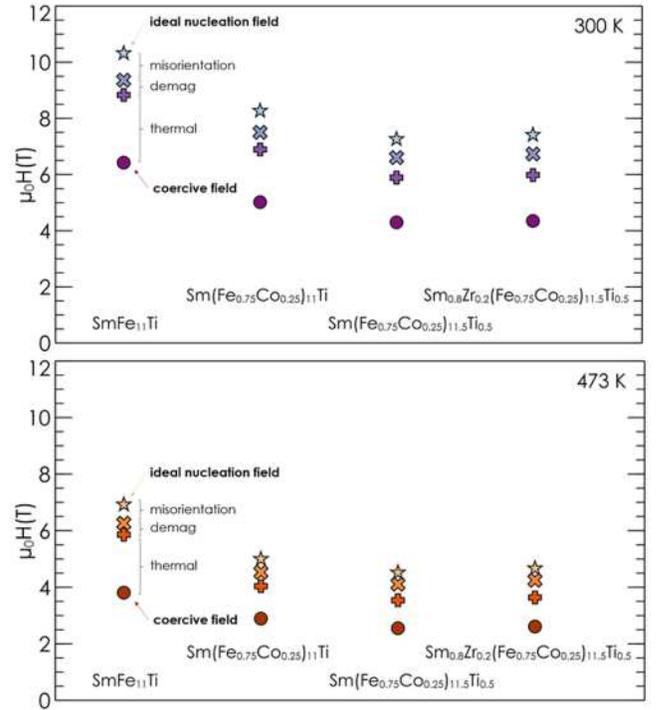}
\caption{\label{fig_limits} Reduction of the ideal nucleation field in various Sm$_{1-z}$Zr$_z$(Fe$_{1-y}$Co$_y$)$_{12-x}$Ti$_x$ compounds at $T = 300$~K and $T = 473$~K of a small magnetic cube without structural defects. The fields indicated by stars are the ideal nucleation field. Symbol $\times$ denotes the field taking into account misorientation. The switching fields computed by Brown’s equation \cite{brown1963micromagnetics} are represented by the symbol +. The circles indicate the critical field at which the energy barrier reaches $25 k_\mathrm{B}T$. All fields were computed for a cube with an edge length of 40 nm.}
\end{figure}

A comparison of the numerical results reveals a striking increase in the activation volume from the small cube to the multigrain system which is mainly caused by the presence of the soft magnetic grain boundary phase. Whereas the small cube is a perfect hard magnetic particle, a 3~nm thick soft magnetic phase separates the grains in the granular magnet. In addition, the demagnetizing field from the neighboring grains is acting on the soft phase where magnetization reversal will be initiated. The soft layer present between hard grains in the multigrain structure makes the spatial variation of the magnetic energy more progressive than in the small cube. Thus, a larger volume (by a factor of 6 in the present case) corresponds to the $25 k_\mathrm{B}T$ energy term provided by thermal activation. As evidenced by eq. (\ref{eq_withHf}) and (\ref{eq_Hc}), the fluctuation field is subsequently reduced by the same factor. By moving from the ideal cube to a realistic structure the activation volume increases and the thermal reduction of coercivity decreases. However, the more realistic structure of the magnet also reduces the intrinsic coercivity $H_0$.

Finally, we computed the limits of coercivity for SmFe-based magnets which are considered as candidates for high performance magnets with a rare earth content smaller than Nd$_2$Fe$_{14}$B. For various Sm$_{1-z}$Zr$_z$(Fe$_{1-y}$Co$_y$)$_{12-x}$Ti$_x$ compounds we computed the effects that reduce the ideal nucleation field towards the maximum possible coercive field. The intrinsic material parameters used for the simulations are listed in Table I. Again, the sample was a cube with an edge length of 40 nm. The field was applied at an angle of one degree. Fig. \ref{fig_limits} shows the ideal nucleation field, the coercive field without demagnetizing effects, the intrinsic coercive field, and the coercivity computed with thermal activation at room temperature and at elevated temperature. At $T = 473$~K the maximum possible expected coercive field for (Sm$_{0.8}$Zr$_{0.2}$)(Fe$_{0.75}$Co$_{0.25}$)$_{11.5}$Ti$_{0.5}$ is $\mu_0 H_\mathrm{c} = 2.61$~T. This can be compared with the computed coercivity limit for Nd$_2$Fe$_{14}$B at $T = 450 K$ which is $\mu_0 H_\mathrm{c}  = 1.88$~T. These limits were computed for a small cubic grain without any soft magnetic defects. Rounding the edges of the cube will improve the coercivity owing to a reduction in the local demagnetizing field near the edges and corners.

Using numerical micromagnetics we computed the effects that reduce the ideal nucleation field of permanent magnets towards the coercive field. We found that even for a magnet with perfect structure, a small cube without surface defects, coercivity is reduced to $60 \% $ at room temperature and $50 \%$ at 473~K of the ideal nucleation field by small misalignment angle (one degree), the demagnetizing field, and thermal activation. In the case of a more realistic grain assembly, the coercive field is reduced by the presence of intergranular defects (represented here by a soft magnetic layer). However, the effect of thermal activation is significantly reduced, as explained above. Therefore, a competition between two antagonistic effects is revealed: as one approaches ideal hard magnetic properties, the drop in coercivity due to defects is reduced but the drop due to thermal activation is increased. In real materials, defects play a major role, whereas coercive field reduction due to thermal activation is of secondary importance at least up to 300 K.

This work was supported by the Austrian Science Fund (FWF): F4112 SFB ViCoM and the pioneering program  $"$Development of magnetic material technology for high-efficiency motors $"$ (2012–) commissioned by the New Energy and Industrial Technology Development Organization (NEDO). 



\bibliography{limitsbib}

\end{document}